\documentclass[5p]{elsarticle}
\usepackage{amsmath,amssymb}
\usepackage{lineno,hyperref}
\modulolinenumbers[5]

\makeatletter
\def\ps@pprintTitle{%
 \let\@oddhead\@empty
 \let\@evenhead\@empty
 \def\@oddfoot{}%
 \def\@oddfoot{\hfill\arabic{page}\hfill}%
 \let\@evenfoot\@oddfoot}
\makeatother

\def\WBR{\discretionary{}{}{}}

\newcommand{\lsim}{\mbox{\raisebox{-0.6ex}{$\stackrel{<}{\sim}$}}\:}
\newcommand{\D}{\mathrm{D}}

\renewcommand{\d}{\mathrm{d}}
\newcommand{\init}{_{\mathrm{i}}}
\newcommand{\Tc}{T_\mathrm{c}}
\newcommand{\eqdef}{=}
\newcommand{\thatis}{\textit{i.e.}}
\newcommand{\tauR}{\tau_{\mathrm{R}}}
\newcommand{\tauRz}{\tau_{\mathrm{R}0}}

\usepackage[normalem]{ulem}  % \sout{old text} for strikeout
\usepackage{color} % For blue in-text comments and additions
\renewcommand\sout{\bgroup \color{red} \ULdepth=-.5ex \ULset}
\begin{document}

\begin{frontmatter}

\title{Hydrodynamic fluctuations of entropy in one-dimensionally expanding system}

\address[1ad]{Department of Physics, Sophia University, Tokyo 102-8554, Japan}
\address[2ad]{Department of Physics, The University of Tokyo, Hongo~7-3-1, Bunkyo-ku, Tokyo 113-0033, Japan}

\author[1ad]{Tetsufumi~Hirano}
\ead{hirano@sophia.ac.jp}

\author[2ad]{Ryuichi~Kurita}

\author[1ad]{Koichi~Murase}
\ead{murase@sophia.ac.jp}

\begin{abstract}
The fluctuation-dissipation relation tells that
dissipation always accompanies with thermal fluctuations.
Relativistic fluctuating hydrodynamics is
used to study the effects of the thermal fluctuations
in the hydrodynamic expansion of the quark-gluon plasma
created in the high-energy nuclear collisions.
We show that the thermal noise obeys the steady-state fluctuation theorem
when (i) the time scales of the evolution of thermodynamic quantities
are sufficiently longer than the relaxation time,
and (ii) the thermal fluctuations of temperature are sufficiently small.
The steady-state fluctuation theorem describes the distribution of the entropy
which can be related to the multiplicity observed in high-energy nuclear collisions.
As a consequence, we propose an upper bound to the multiplicity fluctuations
which is useful to test the initial state models.
We also numerically investigate breaking of the steady-state fluctuation theorem
due to the non-vanishing relaxation time in real nuclear collisions.
\end{abstract}

%\begin{keyword}
%\texttt{elsarticle.cls}\sep \LaTeX\sep Elsevier \sep template
%\MSC[2010] 00-01\sep 99-00
%\end{keyword}

\end{frontmatter}

%\linenumbers
%==============================================================================
\section{Introduction}
Under extremely hot and/or dense circumstance,
quarks and gluons inside hadrons
are deconfined to form quark-gluon plasma (QGP).
The QGP can be created experimentally
in high-energy nuclear collisions
at Relativistic Heavy Ion Collider (RHIC)
in Brookhaven National Laboratory
and at Large Hadron Collider (LHC) in CERN\@.
Just after RHIC started its operation,
relativistic ideal hydrodynamics turned out to work reasonably well for
the description of the spacetime evolution of the QGP~\cite{Kolb:2000sd,Teaney:2000cw,Hirano:2002ds}.
Simulations of relativistic dissipative hydrodynamics
have been extensively performed so far towards further quantitative understanding
of the transport properties of the QGP~\cite{Romatschke:2007mq,Song:2010mg}.
Detailed hydrodynamic studies indicate
the ratio of shear viscosity to entropy density
in the QGP
is very small~\cite{Luzum:2008cw,Song:2008hj,Schenke:2011tv}.

In the past years,
various kinds of fluctuations have attracted a lot of theoretical and experimental
attention.
Observed higher harmonics of azimuthal anisotropy~%
\cite{Adare:2011tg,Sorensen:2011fb,ALICE:2011ab,ATLAS:2012at,Chatrchyan:2012wg}
is explained by initial fluctuations of the transverse profile of the QGP~\cite{Alver:2010gr}.
Hydrodynamic responses of the QGP to the
initial fluctuations give a reasonable interpretation of the higher harmonics.
Another example is the analysis of fluctuations of conserved charges
which could be used to find the signal of quantum chromodynamics (QCD) critical point
where the phase transition from the QGP to hadrons is the second-order one~\cite{Stephanov:1998dy,Stephanov:1999zu}.

The fluctuations to be addressed in this paper are thermal fluctuations
appearing in hydrodynamics.
Thermal equilibrium is a state of maximum entropy in a macroscopic sense.
However, the system is always microscopically fluctuating due to the thermal noises,
\thatis, the thermodynamic variables slightly deviate
from their expectation values on an event-by-event basis in the thermal equilibrium state.
This process reduces the entropy of the system.
At the same time, the system relaxes to the equilibrium state due to the dissipation,
which generates the entropy.
These two processes are compensated with each other
for the system to be stable and
to maintain the entropy around its maximum.
The relation that holds between these two is called
the fluctuation-dissipation relation (FDR).
In the hydrodynamic language,
the dissipative currents, such as the shear stress and the diffusion currents,
are driven by systematic forces (thermodynamic forces)
and, at the same time, random forces (hydrodynamic fluctuations) on an event-by-event basis.
In the hydrodynamic description of high-energy nuclear collision process,
while the dissipative effects such as shear viscosity have been taken into account,
the effects of hydrodynamic fluctuations have not been widely discussed.
However, in a viewpoint of the fluctuation-dissipation theorem,
both effects should be consistently discussed together.
In this paper, employing a relativistic version of second-order fluctuating hydrodynamics~\cite{Murase:2013tma},
we investigate the fluctuations of entropy production
under longitudinally boost-invariant (Bjorken) expansion~\cite{Bjorken:1982qr} in high-energy nuclear collisions.

In many years, the linear response theory~\cite{Kubo:1957mj} has been the milestone
in non-equilibrium statistical mechanics.
On the other hand, in these years, the \textit{fluctuation theorem} \cite{PhysRevLett.71.2401,PhysRevLett.74.2694,PhysRevE.52.5839} has been established
to quantify production of entropy of the system away from
equilibrium. Since the theorem includes
the FDR in the system close to equilibrium,
it is often regarded as a general framework
to analyse the dynamical system even far from equilibrium.
In this paper, we discuss the fluctuation theorem in the context of
the physics of high-energy nuclear collisions for the first time.

The paper is organised as follows:
In Sec.~\ref{sec:rfh},
we briefly introduce relativistic fluctuating hydrodynamics
and apply it to longitudinally boost-invariant expansion.
First we show that
the steady-state fluctuation theorem exactly holds
for the hydrodynamic system under some idealised limits in Sec.~\ref{sec:SSFT}.
In Sec.~\ref{sec:entropy_fluc},
we discuss the consequence in the observables of high-energy nuclear collisions.
In particular
we obtain an upper bound of the power of final-entropy fluctuations
and discuss a consequence to the experimental multiplicities.
Next, in Sec.~\ref{sec:simulation}, we perform numerical simulations of the Bjorken expansion
to quantify the breaking of the steady-state fluctuation theorem
in a realistic condition.
In Sec.~\ref{sec:ssft-breaking},
we discuss two effects breaking the steady-state fluctuation theorem,
\thatis, the finite relaxation time
and the fluctuations of the temperature caused by the hydrodynamic fluctuations.
Finally Sec.~\ref{sec:conclusion} is devoted to the conclusion.

In this paper, we employ natural units, $c = \hbar = k_B = 1$,
and the Minkowski metric, $g_{\mu\nu}=\mathrm{diag}(1,\, -1,\, -1,\, -1)$.

%==============================================================================
\section{Relativistic fluctuating hydrodynamics in Bjorken expansion}
\label{sec:rfh}
In this section we review relativistic fluctuating hydrodynamics
and obtain the expression of the FDR in the Bjorken expansion.

Hydrodynamic equations are the continuity equation for energy and momentum,
\begin{equation}
  \partial_{\mu}T^{\mu \nu} = 0.
  \label{eq:eom}
\end{equation}
The energy-momentum tensor $T^{\mu\nu}$ in the Landau (energy) frame is written down as
\begin{equation}
  T^{\mu\nu} = eu^{\mu}u^{\nu} - (p+\Pi)\Delta^{\mu\nu} + \pi^{\mu\nu},
\end{equation}
where $e$, $p$, $\pi^{\mu\nu}$ and $\Pi$ are
the energy density, the equilibrium pressure, the shear stress tensor and the bulk pressure, respectively.
In this paper we do not consider the other conserved currents.
The tensor $\Delta^{\mu \nu} = g^{\mu \nu} - u^{\mu} u^{\nu}$
is the projector onto the space perpendicular to the four--flow velocity $u^{\mu}$.
To close Eq.~\eqref{eq:eom}, one needs the assumption on the equation of state $p=p(e)$ and the
constitutive equations for the dissipative currents, $\pi^{\mu \nu}$ and $\Pi$.
In relativistic fluctuating hydrodynamics,
the second-order constitutive equations
can be written as stochastic equations~\cite{Murase:2013tma}:
\begin{align}
  \tau_{\pi}\Delta^{\mu\nu}{}_{\alpha\beta} \D\pi^{\alpha\beta} + \pi^{\mu\nu}
    &= 2\eta \Delta^{\mu\nu}{}_{\alpha\beta} \partial^{\alpha}u^{\beta} + \xi^{\mu \nu}, \label{eq:pi} \\
  \left(\tau_{\Pi} \D + 1\right)\Pi
    &= -\zeta \theta + \xi, \label{eq:Pi}
\end{align}
where
transport coefficients $\eta$ ($\zeta$) and $\tau_{\pi}$ ($\tau_{\Pi}$) are the shear (bulk) viscosity
and the relaxation time for the shear stress tensor (bulk pressure), respectively.
The tensor $\Delta^{\mu\nu\alpha\beta}=\frac{1}{2}(\Delta^{\mu\alpha}\Delta^{\nu\beta}+\Delta^{\mu\beta}
\Delta^{\nu\alpha})-\frac{1}{3}\Delta^{\mu\nu}\Delta^{\alpha\beta}$
is a projector for second rank tensors
onto the symmetric and traceless components transverse to the flow velocity.
The operator $\D=u_{\alpha} \partial^{\alpha}$
is the time derivative along the flow velocity, and $\theta = \partial_{\alpha}u^{\alpha}$
is the expansion scalar.
The noise terms $\xi^{\mu\nu}$ and $\xi$
are the hydrodynamic fluctuations of the shear stress and bulk pressure, respectively,
whose intensities are given by the FDR~\cite{Murase:2013tma}:
\begin{align}
  \langle\xi^{\mu \nu}(x)\xi^{\alpha \beta}(x')\rangle
    &= 4T\eta\Delta^{\mu\nu\alpha\beta} \delta^{(4)}(x-x'), \label{eq:ximunu} \\
  \langle\xi(x)\xi(x')\rangle &= 2T\zeta\delta^{(4)}(x-x'), \label{eq:xi}
\end{align}
where $\langle O\rangle$ denotes the average
with respect to the hydrodynamic fluctuations.

In the Bjorken expansion~\cite{Bjorken:1982qr}, the flow velocity is given
by $u^{\mu}=(t/{\tau},0,0,z/{\tau}) =
(\cosh \eta_s, 0, 0, \sinh \eta_s)$,
where $\tau =\sqrt{t^2-z^2}$ and $\eta_s = (1/2)\ln[(t+z)/(t-z)]$
are the proper time and the spacetime rapidity, respectively.
The energy-momentum conservation \eqref{eq:eom}
is reduced to the time evolution of energy density,
\begin{equation}
  \label{eq:evo.e}
  \frac{\d{e}}{\d{\tau}} = -\frac{e+p}{\tau}\left(1-\frac{\pi-\Pi}{sT}\right),
\end{equation}
where $s=(e+p)/T$ is the entropy density and $\pi = \pi^{00}-\pi^{33}$.
From Eqs.~\eqref{eq:pi} and \eqref{eq:Pi}, 
we obtain the following constitutive equations for $\pi$ and $\Pi$:
\begin{align}
  \left(\tau_{\pi}\frac{\d}{\d\tau} + 1\right)\pi &= \frac{4\eta}{3\tau} + \xi_{\pi}, \label{eq:pi_Bj}\\
  \left(\tau_{\Pi}\frac{\d}{\d\tau} + 1\right)\Pi &= -\frac{\zeta}{\tau} + \xi_{\Pi}. \label{eq:Pi_Bj}
\end{align}
Here, to properly define the noise terms,
we need to introduce a ``fluid element'', that we observe,
with the expanding volume of $\tau \Delta\eta_s \Delta x \Delta y$
with $\Delta\eta_s$, $\Delta x$ or $\Delta y$
being the length of the fluid element in each direction.
The noise terms $\xi_{\pi}=\bar\xi^{00} - \bar\xi^{33}$ and $\xi_{\Pi} = \bar\xi$
are hydrodynamic fluctuations for $\pi$ and $\Pi$, where
$\bar\xi^{\mu\nu}$ and $\bar\xi$ denote the volume average of $\xi^{\mu\nu}(x)$ and $\xi(x)$ within the fluid element, respectively.
Note that in general the hydrodynamic fluctuations
arise independently at each spacetime point to induce inhomogeneity,
which would eventually break the boost-invariant expansion of the system,
but here we neglect such effects by assuming
the fluctuation-induced flow is small enough compared to the Bjorken flow.
According to the fluctuation-dissipation relations~\eqref{eq:ximunu}~and~\eqref{eq:xi},
these hydrodynamic noises satisfy
the statistical properties,
\begin{align}
  \langle\xi_{\pi}(\tau)\xi_{\pi}(\tau')\rangle
    &= \frac{8T\eta}{3\tau\Delta\eta_s\Delta x\Delta y} \delta(\tau-\tau'), \label{eq:fdr_shear_BJ} \\
  \langle\xi_{\Pi}(\tau)\xi_{\Pi}(\tau')\rangle
    &= \frac{2T\zeta}{\tau\Delta\eta_s\Delta x\Delta y} \delta(\tau-\tau'). \label{eq:fdr_bulk_BJ}
\end{align}
When the system is close to the local equilibrium,
the noise terms, $\xi_\pi$ and $\xi_\Pi$, have Gaussian distributions.
Once models for the equation of state and
transport coefficients (including relaxation time) are specified,
one can solve the stochastic hydrodynamic equations~\eqref{eq:evo.e}~and~\eqref{eq:pi_Bj},
combined with noises following Eqs.~\eqref{eq:fdr_shear_BJ}~and~\eqref{eq:fdr_bulk_BJ},
as an initial value problem with a given initial condition.
We note that there are two interpretations of the stochastic differential equations,
the It\^{o} integral and the Stratonovich integral.
Here we employ the latter one for the stochastic differential equations in this paper.

%==============================================================================
\section{Steady-state fluctuation theorem in the Bjorken expansion}
\label{sec:SSFT}
% -- Introduction of SSFT --
We first introduce the fluctuation theorem (FT)~\cite{PhysRevLett.71.2401,PhysRevLett.74.2694,PhysRevE.52.5839}.
The FT is a relation
on the distribution of entropy production
in non-equilibrium processes,
which has been proven in various types of systems.
The amount of entropy production
can generally fluctuate from event to event due to thermal fluctuations
even if an initial condition is fixed in a macroscopic sense
since a macroscopic state contains ensemble of different microscopic states.
The steady-state fluctuation theorem (SSFT), which is a certain version of the FT,
gives a relation between two probabilities in stationary processes:
\begin{equation}
  \ln \frac{P(\bar\sigma = \alpha)}{P(\bar\sigma = -\alpha)} = \alpha t,
  \quad (t/t_R \gg 1),
  \label{eq:SSFT}
\end{equation}
where $\sigma$ is entropy production rate, $\bar\sigma$ is its time average,
$t$ is the observation time,
and $t_R$ is the relaxation time scale which characterises the stationary process.
The function $P(\bar\sigma)$ denotes probability density
that the specified average entropy production rate is realised.
Here it should be emphasised that
the entropy can even decrease
at a short time scale at small probability through non-equilibrium processes.
In a strict sense, the SSFT cannot be applied to the Bjorken expansion
because it is not a stationary process.
However, if the expansion time scale is sufficiently longer than the microscopic time scale $t_R$,
we would expect that the SSFT appears as an approximate relation also in the Bjorken expansion.

% -- Note on TFT --
Here we should note that there is a more general version of the FT,
which can be applied to non-stationary processes,
called the transient fluctuation theorem (TFT) which
gives the following relation:
\begin{equation}
  \ln \frac{P(\bar\sigma = \alpha)}{P^\dag(\bar\sigma^\dag = -\alpha)} = \alpha t,
  \label{eq:TFT}
\end{equation}
where $P(\bar\sigma)$ and $P^\dag(\bar\sigma^\dag)$ denote the probability densities
that the specified average entropy production rate is realised
in the considered process
and its corresponding time-reversal process, respectively.
Applying the TFT to the Bjorken expansion,
we might obtain a relation between probabilities of
the Bjorken expansion and its time-reversal process, \thatis, the one-dimensional compression.
However, in high-energy nuclear collisions,
such relativistic compression of thermalised matter cannot be realised experimentally,
and therefore it is difficult to relate the TFT to experimental observables.
For this reason we focus on the SSFT rather than the TFT in this paper.

To discuss the FT in the Bjorken expansion,
we shall first define the entropy production rate
in the fluid element of the volume $\tau\Delta\eta_s \Delta x \Delta y$ as
\begin{align}
  \sigma
    &\eqdef \frac{\d}{\d\tau} (s \tau\Delta\eta_s\Delta x\Delta y) \nonumber \\
    &= \frac{\pi-\Pi}{T}\Delta\eta_s\Delta x\Delta y,
    \label{eq:entropy-production-rate}
\end{align}
where we used Eq.~\eqref{eq:evo.e}
and a thermodynamic relation, $\d{s} = \d{e}/T$,
to obtain the second line.
We note that $s\tau\Delta\eta_s\Delta x \Delta y = (s\gamma)(\tau\Delta\eta_s\Delta x\Delta y / \gamma)$
is a Lorentz-invariant combination with $\gamma = u^0$ being the Lorentz factor.
The average entropy production rate, $\bar\sigma$,
in a time duration from the initial time $\tau\init$ to the current time $\tau$ is written as
\begin{equation}
  \bar\sigma = \frac{1}{\tau-\tau\init}\int_{\tau\init}^{\tau}
    \d\tau' \frac{\pi(\tau')-\Pi(\tau')}{T(\tau')}\Delta\eta_s\Delta x\Delta y,
    \label{eq:average-entropy-production-rate}
\end{equation}
where $\pi(\tau)$ and $\Pi(\tau)$ are formally given
by solving Eqs.~\eqref{eq:pi_Bj} and \eqref{eq:Pi_Bj}:
\begin{align}
  \pi(\tau) &= \int_{\tau\init}^{\tau}\d\tau'
    G_\pi(\tau,\tau')\frac{4\eta(\tau')}{3\tau'} + \delta\pi(\tau),
    \label{pi.i} \\
  \Pi(\tau) &= -\int_{\tau\init}^{\tau}\d\tau'
    G_\Pi(\tau,\tau')\frac{\zeta(\tau')}{\tau'} + \delta\Pi(\tau),
    \label{Pi.i} \\
  G_{\pi/\Pi}(\tau_2,\tau_1) &\eqdef
    \exp\left[-\int_{\tau_1}^{\tau_2}\frac{\d\tau}{\tau_{\pi/\Pi}(\tau)}\right]
    \frac1{\tau_{\pi/\Pi}(\tau_1)}.
    \label{eq:Green}
\end{align}
Here we ignore the terms that depend on
the initial values, $\pi(\tau\init)$ and $\Pi(\tau\init)$,
because these terms damp to vanish
when $\tau-\tau\init \gg \tau_{\pi/\Pi}$.
The time dependence of the transport coefficients,
$\eta(\tau)$, $\zeta(\tau)$ and $\tau_{\pi/\Pi}(\tau)$,
comes from the time evolution of temperature $T(\tau)$.
The fluctuation parts, $\delta\pi(\tau)$ and $\delta\Pi(\tau)$,
are accumulated noises of $\xi_{\pi}$ and $\xi_{\Pi}$, respectively:
\begin{align}
  \delta\pi(\tau)
    &= \int_{\tau\init}^{\tau}\d\tau'
    G_\pi(\tau,\tau')\xi_{\pi}(\tau'),
    \label{eq:fluctuation-part-of-pi} \\
  \delta\Pi(\tau)
    &= \int_{\tau\init}^{\tau}\d\tau'
    G_\Pi(\tau,\tau')\xi_{\Pi}(\tau').
    \label{eq:fluctuation-part-of-bulk}
\end{align}

Next we consider the following two idealised conditions:
(i) the considered volume is sufficiently large
so that the change of the background temperature caused by fluctuations is negligible,
and (ii) the relaxation times, $\tau_\pi$ and $\tau_\Pi$, are sufficiently shorter than
the variation time scale of the temperature and the thermodynamic forces,
\thatis, the Navier--Stokes limit $\tau_{\pi/\Pi} \to 0$ can be safely taken.

Under the condition (ii),
the Green function \eqref{eq:Green}
is reduced to the delta function $\delta(\tau)$,
and the first terms in the right-hand sides
of Eqs.~\eqref{pi.i} and \eqref{Pi.i} become
the Navier--Stokes (first-order) terms.
The integrated noises, $\delta\pi(\tau)$ and $\delta\Pi(\tau)$,
are reduced to $\xi_\pi(\tau)$ and $\xi_\Pi(\tau')$, respectively,
so their correlations are simply given
by Eqs.~\eqref{eq:fdr_shear_BJ}~and~\eqref{eq:fdr_bulk_BJ}.
For the condition (i),
we first define the background $T_0(\tau)$
as the time evolution of the temperature without noises
that is obtained by solving
Eqs.~\eqref{eq:evo.e}--\eqref{eq:Pi_Bj} with $\xi_{\pi}=\xi_{\Pi}=0$.
Using the condition (i),
the temperature and the transport coefficients
in the expression of the entropy production rate
can be replaced by their background values,
$T_0(\tau)$, $\eta_0(\tau) \eqdef \eta(T_0(\tau))$ and $\zeta_0(\tau) \eqdef \zeta(T_0(\tau))$,
to obtain the following expression:
\begin{multline}
  \bar\sigma
    = \frac{\Delta\eta_s\Delta x\Delta y}{\tau-\tau\init}
    \int_{\tau\init}^{\tau}\frac{\d\tau'}{T_0(\tau')} \\
    \times\left[\frac{4\eta_0(\tau')}{3\tau'} + \frac{\zeta_0(\tau')}{\tau'} + \xi_\pi(\tau') - \xi_\Pi(\tau')\right].
  \label{eq:entropy-production-rate-in-navier-stokes-limit}
\end{multline}
In this expression we notice that
the fluctuations contribute to the entropy production
only at the linear order,
and thus the resulting distribution of the entropy production rate
becomes a Gaussian one.
Here we calculate the mean and the variance
that characterise the Gaussian distribution of the entropy production rate:
\begin{align}
  \langle\bar\sigma\rangle
    &= \frac{\Delta\eta_s\Delta x\Delta y}
    {\tau-\tau\init}\int_{\tau\init}^{\tau}\frac{\d\tau'}
    {T_{0}(\tau')}
      \left[\frac{4\eta_0(\tau')}{3\tau'} + \frac{\zeta_0(\tau')}{\tau'}\right], \\
  a^2
    &\eqdef \langle\bar\sigma^2\rangle - \langle\bar\sigma\rangle^2 \nonumber \\
    &= \left(\frac{\Delta\eta_s\Delta x\Delta y}{\tau-\tau\init}\right)^2\int_{\tau\init}^{\tau}
      \d\tau'\int_{\tau_0}^{\tau}\d\tau^{\prime \prime} \nonumber \\
      & \quad\quad\times
      \frac{
        \langle\xi_\pi(\tau')\xi_\pi(\tau'')\rangle +
        \langle\xi_\Pi(\tau')\xi_\Pi(\tau'')\rangle}%
      {T_{0}(\tau')T_{0}(\tau'')} \nonumber \\
    &= \frac{2\Delta\eta_s\Delta x\Delta y}
      {(\tau-\tau\init)^2}\int_{\tau\init}^{\tau}\frac{\d\tau'}
      {T_{0}(\tau')}\left[\frac{4\eta_0(\tau')}{3\tau'}+\frac{\zeta_0(\tau')}{\tau'}\right].
\end{align}

The above expressions of the mean and variance
have the same integral structure,
and in fact we find the following relation:
\begin{equation}
  \frac{2\langle\bar\sigma\rangle}{a^2} = \tau-\tau\init.
  \label{eq:Bj_SSFT}
\end{equation}
Using this relation,
we obtain
a version of the SSFT
in relativistic fluctuating hydrodynamics in the Bjorken expansion:
\begin{align}
  \ln \frac{P(\bar\sigma = \alpha)}{P(\bar\sigma = -\alpha)}
  &= \ln\frac{\exp[-(\alpha - \langle\bar\sigma\rangle)^2/2a^2]}%
      {\exp[-(-\alpha - \langle\bar\sigma\rangle)/2a^2]}
    \label{eq:Bjorken-SSFT-gaussian} \\
  &= \alpha \cdot \frac{2\langle\bar\sigma\rangle}{a^2} \nonumber \\
  &= \alpha\cdot(\tau - \tau_i).
    \label{eq:Bjorken-SSFT}
\end{align}
We note that, in the above expression,
the time duration is measured by the proper time, $\tau$,
unlike the case of the normal SSFT which is measured by the laboratory time, $t$.
This is just because the average entropy production rate~\eqref{eq:average-entropy-production-rate}
is defined by the entropy production per unit proper time,
and there is no essential difference to the normal SSFT\@.

%==============================================================================
\section{Upper bound of entropy fluctuations}
\label{sec:entropy_fluc}
The entropy distribution
can be related to the multiplicity distribution in the high-energy nuclear collisions
since the entropy is approximately proportional to the final multiplicity.
In this section 
we discuss multiplicity fluctuations
through entropy fluctuations in the Bjorken expansion described in the previous section.
In particular, we show the transverse area
dependence of entropy fluctuations and
its upper bound.

To quantify the entropy fluctuations,
we take a ratio of the standard deviation to
the mean value of the entropy distribution at time $\tau$,
where $\tau$ is the time at which entropy becomes no longer produced
due to the freeze-out process of high-energy nuclear collisions.
For one fluid element, the final entropy is $S(\tau) = S\init + \bar\sigma (\tau-\tau_i)$,
and its standard deviation is $\Delta S (\tau) = a (\tau-\tau_i)$.
By taking an ensemble average for a fixed initial entropy
and using Eq.~\eqref{eq:Bj_SSFT}, we obtain the ratio,
\begin{align}
  \frac{\Delta S(\tau)}{\langle S(\tau)\rangle}
    &= \frac{a(\tau-\tau_i)}{S_i + \langle\bar\sigma\rangle(\tau-\tau_i)} \nonumber \\
  	&= \frac{\sqrt{2\langle\bar\sigma\rangle(\tau-\tau_i)}}{S_i + \langle\bar\sigma\rangle(\tau-\tau_i)} \nonumber \\
  	&= \frac{\sqrt{2\langle\delta(\tau s)\rangle}}{\tau\init
      s\init+\langle\delta(\tau s)\rangle}\frac{1}{\sqrt{\Delta\eta_{\mathrm{s}}\Delta x\Delta y}},
\end{align}
where $\delta(\tau s) = \tau s - \tau\init s\init$, and $s\init$ is the initial entropy density.
In the second line we used the fact that the entropy in one fluid element is written as
$S = \tau s\Delta\eta_{\mathrm{s}}\Delta x\Delta y$.
An entire collision system
is considered to be a set of fluid elements,
so we estimate the number of fluid elements in the transverse plane, $n$,
from the transverse area of the system, $A$, as
\begin{equation}
  n = \frac{A}{\Delta x\Delta y}.
\end{equation}
From the assumption of hydrodynamics that
each fluid element can be approximated as a local-equilibrium system,
fluctuations of each fluid element are considered to be statistically independent.
Therefore the relative fluctuations
of the total entropy $\Delta S_\mathrm{tot}$ in the rapidity range $\Delta\eta_s$
are obtained as
\begin{align}
  \frac{\Delta S_{\mathrm{tot}}}{\langle S_{\mathrm{tot}}\rangle}
    &= \frac{1}{\sqrt{n}}\frac{\Delta S(\tau)}{\langle S(\tau)\rangle} \nonumber \\
    &= \frac{\sqrt{2\langle\delta(\tau s)\rangle}}{\tau\init s\init
      +\langle\delta(\tau s)\rangle}\frac{1}{\sqrt{A\Delta\eta_s}}.
  \label{eq:relative-entropy-fluctuation}
\end{align}
The proportionality, $\Delta S_{\mathrm{tot}}/S_{\mathrm{tot}}\propto 1/\sqrt{A\Delta \eta_s}$,
is the common scaling of
the relative fluctuations of macroscopic variables
with respect to the system size.
Here we can identify
$\sqrt{2\langle\delta(\tau s)\rangle}/(\tau\init s\init+\langle\delta(\tau s)\rangle)$
as the constant of the proportionality.
From this proportionality we can say that
the effects of thermal fluctuations are more significant
in smaller systems, such as $p$-A or very peripheral A-A collisions,
if hydrodynamics works in such small systems.

Moreover from Eq.~\eqref{eq:relative-entropy-fluctuation},
we can find an upper bound of the entropy fluctuations:
\begin{align}
  \frac{\Delta S_{\mathrm{tot}}}{\langle S_{\mathrm{tot}}\rangle}
    &\le \frac{1}{\sqrt{2\tau\init s\init}}\frac{1}{\sqrt{A \Delta\eta_s}} = \frac{1}{\sqrt{2S_\mathrm{tot,i}}},
  \label{eq:relative-entropy-fluctuation-upper-bound}
\end{align}
where $S_\mathrm{tot,i} = \tau\init s\init A \Delta\eta_s$ is the initial total entropy in the rapidity range.
To obtain the inequality,
we used a mathematical inequality $\frac{\sqrt{2x}}{a+x} \le \frac{1}{\sqrt{2a}}$
for $x\ge0$ and any positive constant $a$,
where the equality is satisfied in the case $x=a$.
The most important point is that
the upper bound of the entropy fluctuations is solely determined
by the initial total entropy $S_\mathrm{tot,i}$
and does not depend on the details of the intermediate dynamics
such as the equation of state and the value of transport coefficients.
We also note that the upper bound
is independent of the fluid element size, $\tau\Delta\eta_s\Delta x\Delta y$,
which we assumed in the derivation.

Equation~\eqref{eq:relative-entropy-fluctuation-upper-bound}
is the inequality for a fixed initial condition with the total entropy $S_\mathrm{tot,i}$.
To relate the Eq.~\eqref{eq:relative-entropy-fluctuation-upper-bound}
to the experimental multiplicities,
we need to consider event averages over initial conditions.
Here we make two simplifications that
both of $S_\text{tot,i}$ and $\langle S_\text{tot}\rangle$
are proportional to the transverse area $A$,
and that the multiplicity distribution for a fixed final entropy $S_\text{tot}$
follows the Poisson distribution.
Using these simplifications
we can find an upper bound of the multiplicity fluctuations
as follows (see \ref{sec:appendix-multiplicity-fluctuations} for the derivation):
\def\EventAverage#1{\langle#1\rangle_\text{ev}}
\def\SInit{S_\mathrm{tot,i}}
\begin{align}
  \frac{(\Delta_\text{ev} N)^2 - \EventAverage{N}}{\EventAverage{N}^2}
  &\le \frac{(\Delta_\text{ev} \SInit)^2}{\EventAverage{\SInit}^2}
  +\frac1{2\EventAverage{\SInit}},
  \label{eq:multiplicity-fluctuations-upper-bound}
\end{align}
where $N$ is the multiplicity in a considered rapidity range,
$\EventAverage{N}$ and $\EventAverage{\SInit}$
are the event averages of $N$ and the initial entropy $\SInit$,
and $(\Delta_\text{ev} N)^2 = \EventAverage{(N - \EventAverage{N})^2}$
and $(\Delta_\text{ev} \SInit)^2 = \EventAverage{(\SInit - \EventAverage{\SInit})^2}$
are the variance of $N$ and $\SInit$, respectively.
The second term in the left-hand side comes from the Poisson statistics.
Here one notices that the right-hand side is totally written by
the quantities specific to initial conditions,
and the left-hand side can be measured in experiments.
Therefore this inequality may be used
to test initial state models in comparison with experimental data
without relying on any specific modeling of intermediate dynamics.

%=============================================================================
\section{Numerical tests}
\label{sec:simulation}
In Secs.~\ref{sec:SSFT}~and~\ref{sec:entropy_fluc},
we assumed the Navier--Stokes limit where the relaxation time is negligible.
A non-vanishing relaxation time is, however, needed to maintain the causality
in relativistic dissipative hydrodynamics~\cite{Hiscock:1983zz,Hiscock:1985zz}.
In particular, in high-energy nuclear collisions,
the relaxation time is comparable to
the time scale of the evolution of thermodynamic quantities.
In this section, we consider
the effects of the non-vanishing relaxation time on the entropy fluctuations
by solving the stochastic differential equation numerically.

%------------------------------------------------------------------------------
% \subsection{Setup}

\begin{figure}[htb]
  \centering
  \includegraphics[width = 0.9\columnwidth]{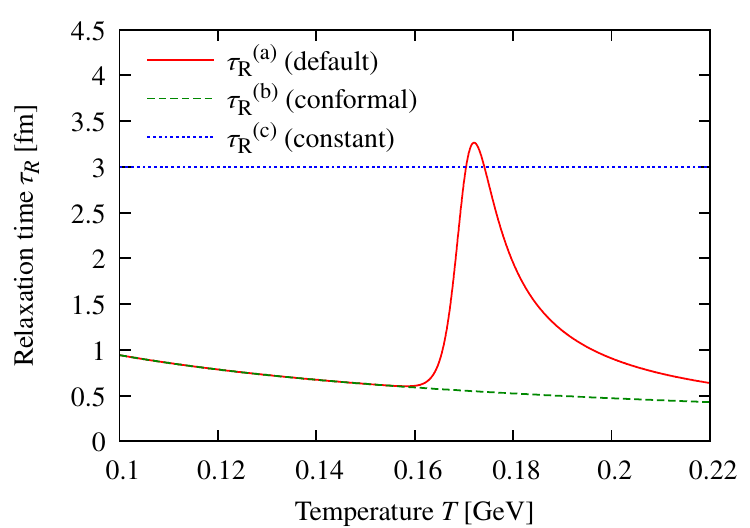}
  \caption{Three models of the relaxation times are shown as functions of temperature.
  The default relaxation time is shown by the red solid line.
  The conformal and constant relaxation times are shown by the green dashed and blue dotted lines.}
  \label{fig:relaxation-time}
\end{figure}

First we define a parametrised equation of state and transport coefficients.
The lattice QCD simulations indicate
a crossover from the hadronic matter to the QGP
on the temperature axis
with vanishing baryon chemical potential~\cite{Aoki:2005vt,Aoki:2006we,Bazavov:2009zn}.
In this study
we employ a model equation of state~\cite{Asakawa:1995zu}
with a crossover,
where the entropy density
as a function of temperature is parametrised as
\begin{align}
  s(T)
    &= \frac{4\pi^2}{90}g_h\;T^3\frac{1-\tanh\left(\frac{T-\Tc}{d}\right)}{2} \nonumber \\
    &+ \frac{4\pi^2}{90}g_q\;T^3\frac{1+\tanh\left(\frac{T-\Tc}{d}\right)}{2}.
\end{align}
Here $g_{\mathrm{h}}=3$ and $g_{\mathrm{q}}=37$ are degrees of freedom of hadrons
and QGP, respectively, in $N_{\mathrm{c}} = 3$ and $N_{\mathrm{f}} = 2$ case.
In this parametrised form,
we can change the crossover temperature $\Tc$ and the crossover region size $d$
to see their effects,
yet for the present study we fix $\Tc = 170\ \text{MeV}$ and $d = \Tc/50$.
Contrary to the equation of state,
less known are the transport coefficients
of the QGP\@.
Hence, just for the purpose of demonstrating
relativistic fluctuating hydrodynamics,
we employ the following parametrisation for transport coefficients
for the shear and bulk viscosity \cite{Policastro:2001yc,Weinberg:1971mx}:
\begin{align}
  \frac{\eta}{s}  &= \frac{1}{4\pi}, \label{eq:shear-viscosity} \\
  \frac{\zeta}{s} &= 15\left(\frac{1}{3}-c_{\mathrm{s}}^2\right)^2\frac{\eta}{s},
\end{align}
where $c_{\mathrm{s}}^2= \d p/\d e$ is the squared sound velocity.
We assume a common relaxation time
for the bulk pressure and the shear stress: $\tauR \eqdef \tau_\pi = \tau_\Pi$.
We mainly consider the relaxation time given by Refs.~\cite{Grad1949,Muller1967,ISRAEL1979341}: %ISRAEL1976310,Stewart59
\begin{equation}
  \tauR^\text{(a)} = \frac{3\eta}{2p}.
  \label{eq:default-relaxation-time}
\end{equation}
For the purpose of investigating the effects of the relaxation time on the SSFT,
we also consider other models of relaxation times.
One is the conformal one, $\tauR^\text{(b)} = 3/2\pi T$, obtained by
applying the equation of state of massless ideal gases
to Eq.~\eqref{eq:default-relaxation-time} along with Eq.~\eqref{eq:shear-viscosity}.
Another one is a constant relaxation time, $\tauR^\text{(c)} = 3.0 \ \text{fm}$.
Figure~\ref{fig:relaxation-time} shows the temperature dependence of the relaxation times.
In the following discussions, the default relaxation time $\tauR^\text{(a)}$ is used
if it is not explicitly specified.
We choose the initial time $\tau\init = 1.0\ \text{fm}$
and the initial temperature $T\init = 0.22\ \text{GeV}$.
The initial values of the dissipative currents are taken to vanish.
The size of the fluid element is $\Delta\eta_s = 1$ and $\Delta x=\Delta y = 1\ \text{fm}$.
For time integration,
we use the second-order stochastic Runge--Kutta method with the strong order 1 (See \ref{sec:srk2})
based on the improved Euler method,
for which we choose the time step $\Delta\tau = 0.1\ \text{fm}$.

%------------------------------------------------------------------------------
% \subsection{Time evolution}

\begin{figure}[htb]
  \centering
  \includegraphics[width=0.9\columnwidth]{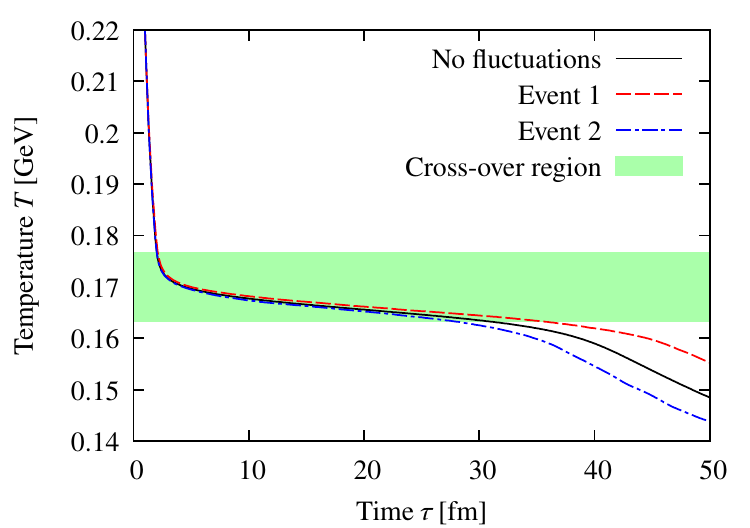}
  \caption{The time evolution of temperature is shown.
  The black solid line shows
  the result of dissipative hydrodynamics without fluctuations.
  The red dashed and blue chain lines are two examples of the results of
  fluctuating hydrodynamics.
  The pale green band shows a crossover region $T \in [\Tc - 2d, \Tc + 2d]$.}
  \label{fig:temperature-vs-time}
\end{figure}

\begin{figure}[htb]
  \centering
  \includegraphics[width=0.9\columnwidth]{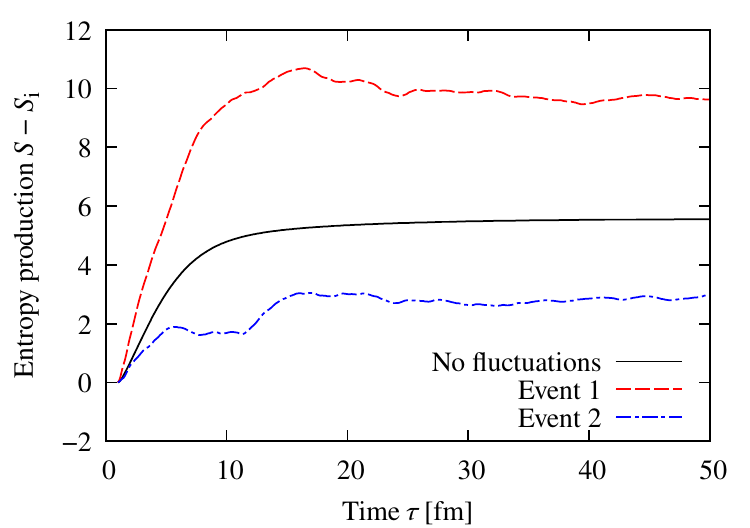}
  \caption{The entropy production in one fluid element,
  $S - S\init = \delta(\tau s)\Delta\eta_s\Delta x\Delta y$,
  is shown. The black solid line shows
  the result of dissipative hydrodynamics without fluctuations.
  The red dashed and blue chain lines are
  the two examples of fluctuating hydrodynamics
  corresponding to those of Fig.~\ref{fig:temperature-vs-time}.}
  \label{fig:entropy}
\end{figure}

Figure~\ref{fig:temperature-vs-time} shows the time evolution of the temperature
in dissipative hydrodynamics without fluctuations
and two sample events from fluctuating hydrodynamics.
One can see that the temperature initially decreases rapidly,
goes down slowly in the crossover region,
and finally decreases rapidly again
after passing through the crossover region
at $\tau \sim 40\text{--}50\ \text{fm}$.
Also the temperature difference
between dissipative hydrodynamics
and fluctuating hydrodynamics
is still an order of a few percent
after the crossover region $\tau \sim 50\ \text{fm}$
in typical events as seen in Fig.~\ref{fig:temperature-vs-time}.
This means that
the condition (i) in Sec.~\ref{sec:SSFT}
is a good approximation in the current setup.

The entropy production for the fluid element
as a function of time is shown in Fig.~\ref{fig:entropy}.
While the total entropy is conserved in ideal hydrodynamics,
it monotonically increases in dissipative hydrodynamics due to the second law of thermodynamics as shown by the solid line.
Specifically, the increase is rapid in the early time ($\tau \lsim 10\ \text{fm}$) and slows down afterwards
since the dissipative currents appearing in the expression of entropy production~\eqref{eq:entropy-production-rate}
are on average proportional to the thermodynamic force which is $1/\tau$ in the Bjorken expansion.
We also show two examples of the entropy production
of fluctuating hydrodynamics
with dashed and chain lines,
which fluctuate around the results of dissipative hydrodynamics.
One notices that the fluctuations of the entropy production
are more significant compared to the temperature fluctuations (shown in Fig.~\ref{fig:temperature-vs-time})
although the entropy is a function of temperature.
This is because the tiny fluctuations of temperature
are magnified by the steep change of the entropy in the crossover region.
One also observes that entropy can even decrease in short time scales,
which can be explained by the fluctuation theorem~\eqref{eq:SSFT}
claiming that the entropy of a small system can decrease in a short time scale
with a small, but still non-zero, probability.
We note that the second law of thermodynamics
corresponds to the fact that the entropy increases on average in fluctuating hydrodynamics.

\begin{figure}[htb]
  \centering
  \includegraphics[width=0.9\columnwidth]{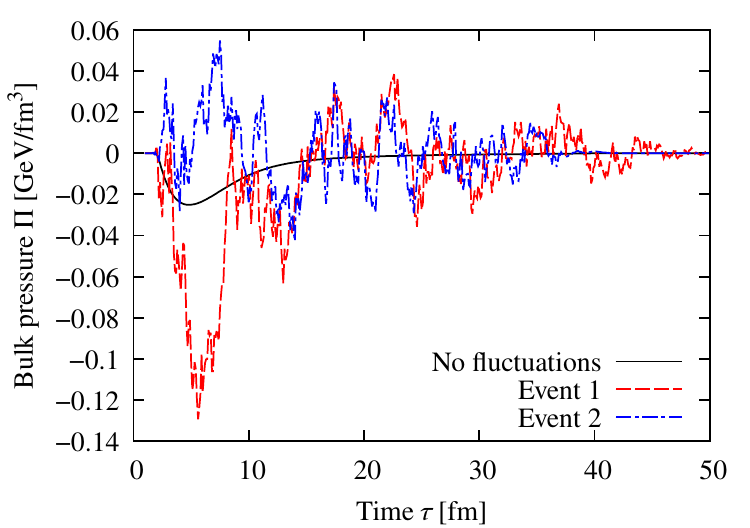} \\
  \includegraphics[width=0.9\columnwidth]{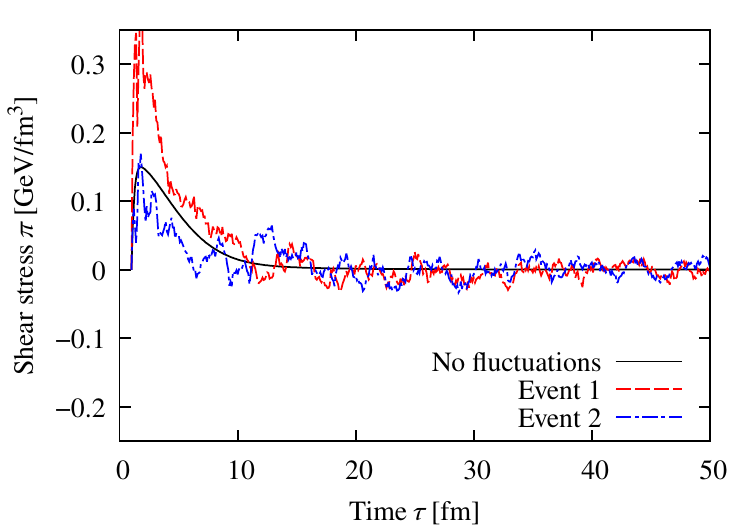}
  \caption{The time evolution of bulk pressure and shear stress are shown
  in the upper and lower panel, respectively.
  The results of dissipative hydrodynamics is shown by the black solid lines.
  The results of fluctuating hydrodynamics is shown by red dashed and blue chain lines
  each of which corresponds to those in Fig.~\ref{fig:temperature-vs-time}.}
  \label{fig:bulk-and-shear}
\end{figure}

Using Eq.~\eqref{eq:entropy-production-rate},
the temporal decrease of the entropy can be attributed
to the behaviours of the bulk pressure $\Pi$ and shear stress $\pi$,
which are shown in Fig.~\ref{fig:bulk-and-shear}.
These dissipative currents of fluctuating hydrodynamics
fluctuate around the ones of dissipative hydrodynamics.
We see that, unlike in expanding systems in dissipative hydrodynamics,
the bulk pressure (shear stress) can be positive (negative) in fluctuating hydrodynamics,
which causes the negative entropy production rate,
$\sigma \propto \pi - \Pi < 0$.

%------------------------------------------------------------------------------
% \subsection{SSFT}

\begin{figure}[htb]
  \includegraphics[width=0.9\columnwidth]{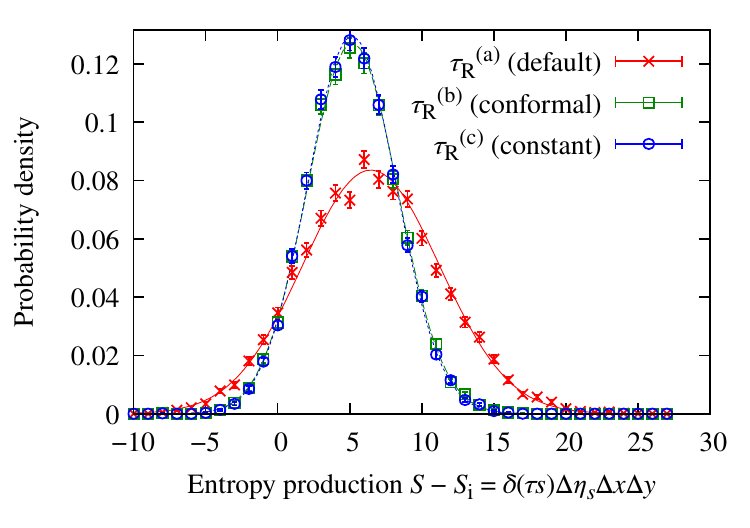}
  \caption{Probability distributions of entropy production in one fluid element
  at $\tau=50\ \text{fm}$ are shown for the three models of relaxation times.
  The error bars show statistical errors.
  The curves are fitted by Gaussian distributions.}
  \label{fig:hist}
\end{figure}

So far we have discussed the time evolution of fluctuating hydrodynamics using two sample events.
Now, to discuss the SSFT in the Bjorken expansion,
we perform 10000 events of simulations for each model of the relaxation time
and obtain the probability distribution of the entropy production.
Figure~\ref{fig:hist} shows the entropy production distribution
in the fluid element at $\tau = 50\ \text{fm}$
for the three models of the relaxation time.
One can observe that there are non-negligible probabilities
that the entropy production becomes negative,
\thatis, the final entropy becomes smaller than the initial value.
The probabilities of negative entropy production
are 8.42(28)\%, 3.41(18)\% and 3.70(18)\%
for $\tauR^\text{(a)}$, $\tauR^\text{(b)}$ and $\tauR^\text{(c)}$, respectively.
One can see that for all the relaxation time
the distribution is well fitted by Gaussian
although each can have quite different mean and variance.
This means that
the equality~\eqref{eq:Bjorken-SSFT-gaussian}
in the derivation of the SSFT~\eqref{eq:Bjorken-SSFT}
is still valid for the current numerical setup,
so the relation~\eqref{eq:Bj_SSFT} is
the remaining key to check the SSFT\@.

\begin{figure}[htb]
  \includegraphics[width=0.9\columnwidth]{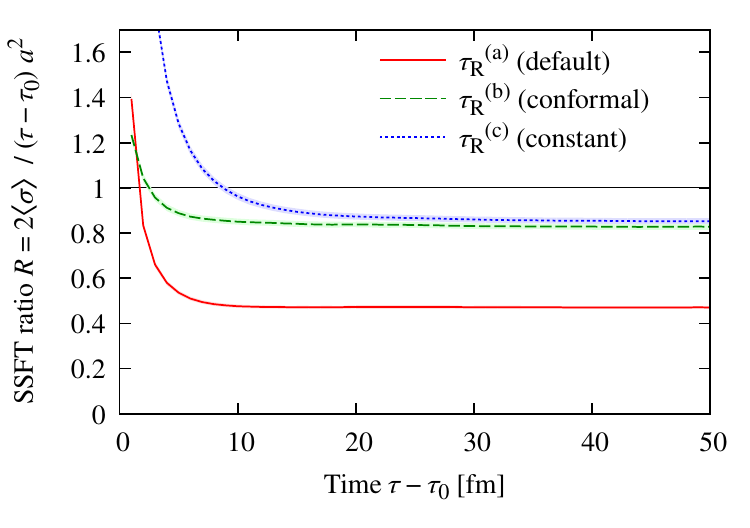}
  \caption{The SSFT ratio $R$ is shown as a function of the time.
  Statistical errors are shown by bands.}
  \label{fig:ssft-ratio-vs-time}
\end{figure}
To see if the relation~\eqref{eq:Bj_SSFT} holds for the current setup,
it is useful to calculate the following ratio:
\begin{align}
  R &\eqdef \frac{2\langle\bar\sigma\rangle}{a^2\cdot(\tau - \tau\init)}.
  \label{eq:definition-of-R}
\end{align}
Here $R=1$ means the SSFT,
and its deviation from unity measures the breaking of the SSFT\@.
The time dependence of the ratio $R$
is shown in Fig.~\ref{fig:ssft-ratio-vs-time}
for each model of the relaxation time.
The ratio has very small values
at the initial stage, $\tau - \tau\init \lsim \tauR$,
and then converges to a final value at the later stage, $\tau - \tau\init \gg \tauR$,
which is consistent with the SSFT~\eqref{eq:SSFT}.
The ratio for the default relaxation time becomes finally $R \sim 0.56$
which is significantly smaller than unity.
To study what breaks the SSFT,
the result can be compared to those of the other relaxation times:
Both of the ratios for the other two relaxation times
successfully converge to values close to unity,
which means that the SSFT is approximately valid for these relaxation times.
The differences of the default relaxation time and these relaxation times
lie in the temperature dependence: The default relaxation time peaks
around the crossover temperature.
The rapid variation of the relaxation time caused by this strong temperature dependence
breaks the SSFT as we will see in the next section.
This means that the SSFT ratio $R$ is sensitive
to the temperature dependence of the relaxation time.

%==============================================================================
\section{SSFT breaking}
\label{sec:ssft-breaking}
In the previous section
we observed breaking of the SSFT by numerical simulations.
In this section, to get a deeper understanding,
we discuss two effects that break the SSFT
each of which corresponds to an idealised condition introduced in Sec.~\ref{sec:SSFT},
\thatis, the effects of
(i) background fluctuations caused by the hydrodynamic fluctuations
and (ii) non-vanishing relaxation time.

%------------------------------------------------------------------------------
\subsection{Background fluctuations}
\begin{figure}[htb]
  \includegraphics[width=0.9\columnwidth]{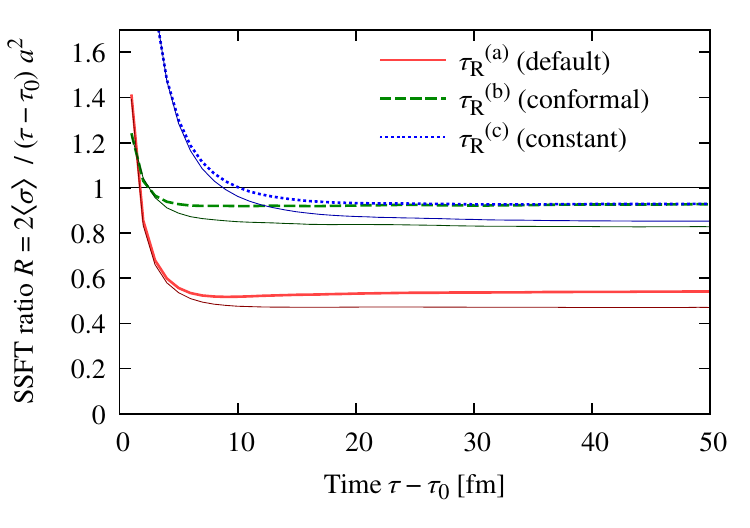}
  \caption{The SSFT ratios calculated under the non-fluctuating background $T_0(\tau)$
    are compared to those of the full non-linear time evolution shown in Fig.~\ref{fig:ssft-ratio-vs-time}.
    The thick lines and thin solid lines
    correspond to the SSFT ratios under the non-fluctuating background and in the full time evolution, respectively.
    The red solid, green dashed and blue dotted lines
    represent the results for the three relaxation models, respectively.}
  \label{fig:ssft-ratio-background}
\end{figure}
Here we discuss the effects of temperature fluctuations numerically.
We first obtain the evolution of non-fluctuating background temperature, $T_0(\tau)$,
by numerically solving Eqs.~\eqref{eq:evo.e}--\eqref{eq:Pi_Bj} without hydrodynamic fluctuations.
Then we solve the constitutive equations~%
\eqref{eq:pi_Bj} and \eqref{eq:Pi_Bj}
under the non-fluctuating background $T_0(\tau)$
and finally perform event-by-event integrations of the entropy production~%
\eqref{eq:average-entropy-production-rate}.
In Fig.~\ref{fig:ssft-ratio-background},
the time evolution of the SSFT ratio under the non-fluctuating background
is compared to the full non-linear time evolution
for each relaxation time model.
One can observe that the corrections
due to the fluctuating background are around $\sim 0.1$
and have weak dependence on the relaxation time models.

%------------------------------------------------------------------------------
\subsection{Non-vanishing relaxation time}
We next discuss effects of non-vanishing relaxation time
under the non-fluctuating background evolution $T_0(\tau)$.
We now consider the case $\tau_\pi = \tau_\Pi = \tauR$
following our numerical setup in the previous section.
Combining Eqs.~\eqref{eq:average-entropy-production-rate}--\eqref{eq:fluctuation-part-of-bulk},
the expression of entropy production is written as
\def\cellVolume{\Delta V}
\begin{align}
  \bar\sigma
  &= \frac{\cellVolume}{\tau - \tau\init}
    \int_{\tau\init}^\tau \frac{\d\tau_1}{T_0(\tau_1)} 
    \int_{\tau\init}^{\tau_1} \d\tau_2 G_0(\tau_1, \tau_2) F(\tau_2) \nonumber \\
  &= \frac{\cellVolume}{\tau - \tau\init}
    \int_{\tau\init}^{\tau} \d\tau_2
    \int_{\tau_2}^{\tau} \frac{\d\tau_1}{T_0(\tau_1)}
    G_0(\tau_1, \tau_2) F(\tau_2),
    \label{eq:average-entropy-production-ratio.full-commute}
\end{align}
where $\Delta V \eqdef \Delta\eta_s\Delta x\Delta y$,
$F(\tau) \eqdef 3\eta_0(\tau)/4\tau + \zeta_0(\tau)/\tau + \xi_\pi(\tau) - \xi_\Pi(\tau)$,
and the integral kernel $G_0(\tau_1, \tau_2)$ is defined
using the background relaxation time $\tauRz(\tau)\WBR\ \eqdef\ \WBR\tauR(T_0(\tau))$.
We then use the relation $[1+\tauRz(\tau_1) \d/\d\tau_1] \WBR G_0(\tau_1, \tau_2) \WBR = \WBR 0$
to replace $G_0(\tau_1, \tau_2)$ with $-\tauRz(\tau_1) (\d/\d\tau_1) \WBR G_0(\tau_1, \tau_2)$
and perform integration by parts with respect to $\tau_1$
to obtain the following expression:
\begin{multline}
  \bar\sigma = \frac{\cellVolume}{\tau - \tau\init}\biggl[
    \int_{\tau\init}^{\tau} \d\tau_2\frac{F(\tau_2)}{T_0(\tau_2)} \\
    -\frac{\tauRz(\tau)}{T_0(\tau)} \int_{\tau\init}^\tau \d\tau_2 G_0(\tau, \tau_2) F(\tau_2) \\
    +\int_{\tau\init}^{\tau} \d\tau_1 \Bigl(\frac{\d}{\d\tau_1}\frac{\tauRz(\tau_1)}{T_0(\tau_1)}\Bigr)
    \int_{\tau\init}^{\tau_1} \d\tau_2 G_0(\tau_1, \tau_2) F(\tau_2)\biggr],
  \label{eq:entropy-production-rate-in-non-vanishing-relaxation-time}
\end{multline}
where the first and second terms are obtained as surface terms.
One notices that the first term is
exactly the same as the expression in the Navier--Stokes limit~%
\eqref{eq:entropy-production-rate-in-navier-stokes-limit}
while the second and third terms are the corrections
by the non-vanishing relaxation time.
Specifically one can observe that the second and third terms have factors
$(\tauR/T)$ and $\D(\tauR/T)$, respectively,
and therefore they are identified to be the corrections due to
the absolute value and the temporal change of the relaxation time, respectively.

The mean of entropy production is
\begin{align}
  \langle\bar\sigma\rangle
  &= \frac{\cellVolume}{\tau - \tau\init}
    \int_{\tau\init}^\tau \frac{\d\tau_1}{T_0(\tau_1)} 
    \int_{\tau\init}^{\tau_1} \d\tau_2 G_0(\tau_1, \tau_2) \langle F(\tau_2) \rangle,
    \label{eq:finite-relaxation.mean-entropy-production}
\end{align}
where $\langle F(\tau)\rangle = 4\eta_0(\tau)/3\tau + \zeta_0(\tau)/\tau$.
The variance can be calculated
by applying the FDR,
$\langle\delta F(\tau_1)\delta F(\tau_2)\rangle = 2T_0(\tau_1)\langle F(\tau_1)\rangle\delta(\tau_1-\tau_2)/\Delta V$ with $\delta F \eqdef F - \langle F\rangle$,
to the product of
Eqs.~\eqref{eq:average-entropy-production-ratio.full-commute}
and \eqref{eq:entropy-production-rate-in-non-vanishing-relaxation-time}.
As a result we obtain three terms each of which corresponds to each term in
Eq.~\eqref{eq:entropy-production-rate-in-non-vanishing-relaxation-time}:
\begin{align}
  a^2 &= 2(\langle\bar\sigma\rangle + \gamma + \delta).
  \label{eq:entropy-production-variance-3-terms}
\end{align}
Here
\begin{align}
  \gamma &=-\frac{\cellVolume}{\tau-\tau\init} \frac{\tauRz(\tau)}{T(\tau)}
    \int_{\tau\init}^{\tau} \frac{\d\tau_1}{T_0(\tau_1)}
  \nonumber \\ & \times
    \int_{\tau\init}^{\tau_1} \d\tau_2
    G_0(\tau, \tau_2) G_0(\tau_1, \tau_2) T_0(\tau_2)\langle F(\tau_2)\rangle,
    \label{eq:finite-relaxation.gamma-correction} \\
  \delta &= \frac{\cellVolume}{\tau-\tau\init}
    \int_{\tau\init}^{\tau} \frac{\d\tau_1}{T_0(\tau_1)}
    \int_{\tau\init}^{\tau} \frac{\d\tau_3}{T_0(\tau_3)}
    \left[T_0(\tau_3)\D_3\frac{\tauRz(\tau_3)}{T_0(\tau_3)}\right]
  \nonumber \\ & \times
    \int_{\tau\init}^{\tau_{\min}} \d\tau_2
    G_0(\tau_1, \tau_2) G_0(\tau_3, \tau_2) T_0(\tau_2)\langle F(\tau_2)\rangle,
    \label{eq:finite-relaxation.delta-correction}
\end{align}
where $\D_3 \eqdef \d/\d\tau_3$ and $\tau_{\min} \eqdef \min\{\tau_1, \tau_3\}$.

Now we discuss the deviation of the SSFT ratio, $R$, from unity.
The inverse ratio is obtained from Eq.~\eqref{eq:entropy-production-variance-3-terms}:
\begin{align}
  R^{-1} &= 1
    + \frac{\gamma}{\langle\bar\sigma\rangle}
    + \frac{\delta}{\langle\bar\sigma\rangle}.
\end{align}
First we discuss the second term $\gamma$.
One can observe in Eq.~\eqref{eq:finite-relaxation.gamma-correction}
that it contains an explicit $\tauRz$ dependence only outside of the integration,
and therefore the term is considered to be the correction to $R$
solely due to the finiteness of the relaxation time.
The third term $\delta$ is considered to be the correction due to
the temporal change of the relaxation time
as one can see that it is proportional to a dimensionless factor $T \D(\tauR/T)$
in Eq.~\eqref{eq:finite-relaxation.delta-correction}.
From magnitudes of the two terms in $T \D(\tauR/T) = \D \tauR - (\tauR/T) \D T$,
the conditions for the vanishing third term read:
\begin{align}
  \D \tauR &\ll 1, \\
  \tauR &\ll 1/(\D \ln T).
\end{align}
The second condition implies that the relaxation time should
be sufficiently shorter than the hydrodynamic time scale of temperature change.
The first condition can be interpreted similarly:
The relaxation time should be shorter than the variation time scale of the relaxation time itself
since the condition can be rewritten as $\tauR \ll 1/(\D\ln\tauR)$.
The large deviation of $R$ from unity with the relaxation time model $\tauR^{(a)}$ in Fig.~\ref{fig:ssft-ratio-vs-time}
can be understood by this effect. The relaxation time $\tauR^{(a)}(T)$ have a peak structure
in its temperature dependence, so it changes rapidly in the time evolution to break the condition $\D\tauR \ll 1$.
For the other relaxation time models, the temporal change of $\tauR$ is milder,
which explains their smaller deviation of $R$ from unity.

\begin{figure}[htb]
  \includegraphics[width=0.9\columnwidth]{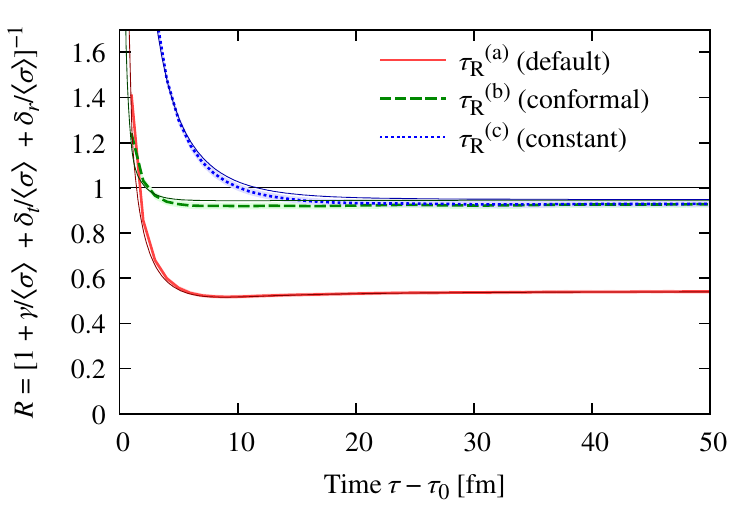}
  \caption{
    The SSFT ratios evaluated by numerical integrations of
    the analytic expressions~\eqref{eq:finite-relaxation.mean-entropy-production},
    \eqref{eq:finite-relaxation.gamma-correction} and
    \eqref{eq:finite-relaxation.delta-correction} are
    compared to those by the numerical simulations under the non-fluctuating background $T_0(\tau)$
    shown in Fig.~\ref{fig:ssft-ratio-integral}.
    The thin solid lines and thick lines
    represent the results for the analytic expressions and the event-by-event numerical simulations, respectively.
    The red, green and blue lines correspond to the results for the three relaxation models, respectively.
    The bands with pale colours show the statistical errors of the results of the numerical simulations.}
  \label{fig:ssft-ratio-integral}
\end{figure}
\begin{figure}[htb]
  \includegraphics[width=0.9\columnwidth]{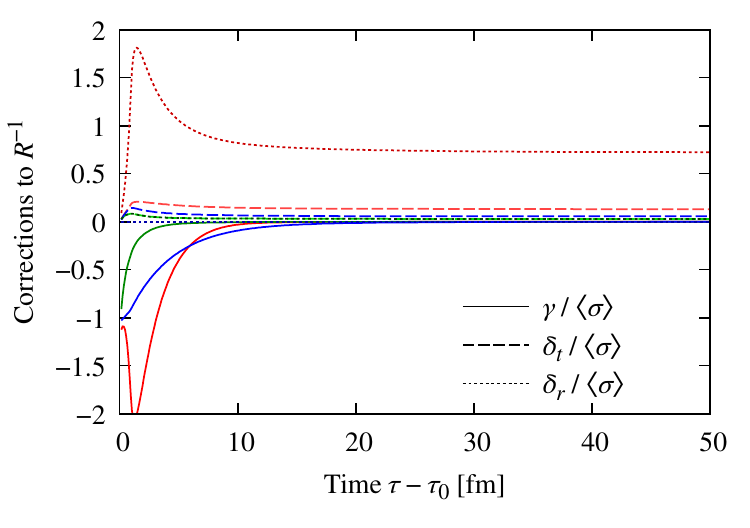}
  \caption{
    The corrections to the SSFT factor $R$ are shown.
    The solid, dashed and dotted line represent
    the corrections due to the finiteness of the relaxation time $\gamma$,
    the temperature evolution $\delta_t$ and the relaxation time evolution $\delta_r$, respectively.
    The red, green and blue lines correspond to the results for the three relaxation models,
    respectively, as the same as the other figures.}
  \label{fig:ssft-corrections}
\end{figure}

To quantify the effects of each correction
we perform numerical integrations of these analytic expressions~%
\eqref{eq:finite-relaxation.mean-entropy-production},
\eqref{eq:finite-relaxation.gamma-correction} and
\eqref{eq:finite-relaxation.delta-correction}.
In particular we separate
two different contributions from $\delta$ corrections, $\delta = \delta_t + \delta_r$,
where the temperature evolution effect $\delta_t$
and the relaxation time evolution effect $\delta_r$ correspond
to the two terms in $T\D(\tauR/T) = -(\tauR/T^2)\D T + \D\tauR$, respectively.
For an efficient evaluation of the time dependence of the integrations,
we construct dynamical equations of the integrals (See \ref{sec:numerical-integration}).
Figure~\ref{fig:ssft-ratio-integral}
shows the time dependence of the SSFT ratio evaluated by numerical integration of the analytic expressions
compared to the results by the event-by-event numerical simulations
under the non-fluctuating background $T_0(\tau)$.
One can see that the results of the event-by-event numerical simulations
are reproduced by the analytic expressions within statistical errors.

Figure~\ref{fig:ssft-corrections}
shows the time dependence of each correction to the SSFT factor $R$.
One can see that the corrections $\gamma$ shown by solid lines
vanish within $10$--$20\ \text{fm}$,
and therefore the remaining contributions at the final time
are only $\delta_t$ and $\delta_r$.
The corrections $\delta_t$ shown by dashed lines
have relatively the same order of contributions
while the corrections $\delta_r$ shown by dotted lines
have dramatic differences among the relaxation time models.
The $\delta_r$ correction for the constant relaxation time model
shown by blue dotted line trivially vanishes since
the relaxation time does not change in this model.
The $\delta_r$ correction for the conformal relaxation time model
shown by green dotted line has exactly the same value
with the $\delta_t$ corrections
since in this model the two terms from $\D(\tauR/T) = (1/T)\D\tauR + \tauR\D(1/T)$
give the identical contributions as $\tauR \propto 1/T$.
The $\delta_r$ correction for the default model
shown by red dotted line gives a large correction
because of the fast evolution of the relaxation time
explained by the steep temperature dependence
of the relaxation time around the crossover region.

Thus the significant deviation of the SSFT ratio $R$ of the default relaxation model
in Fig.~\ref{fig:ssft-ratio-vs-time} was quantitatively confirmed to be the $\delta_r$ correction,
the correction due to the time evolution of the relaxation time.
The remaining deviations are explained by the time evolution of the temperature
and the background fluctuations.

%==============================================================================
\section{Summary and concluding remarks}
\label{sec:conclusion}
In this paper we focused on the distribution of entropy production
caused by the hydrodynamic fluctuations, \thatis, the thermal fluctuations of hydrodynamics,
in a simple setup of the Bjorken flow which is a one-dimensionally expanding system.
The dynamics of the system is described by the relativistic fluctuating hydrodynamics
whose equations become stochastic differential equations
due to the noise terms.
In (i) a limit that the considered fluid element is large enough that the background fluctuations are negligible
and (ii) the Navier--Stokes limit where the relaxation time is sufficiently shorter than the time scale of macroscopic dynamics of hydrodynamic fields,
we have shown a relation in the expanding system~\eqref{eq:Bj_SSFT} which shares the same structure with the SSFT~\eqref{eq:SSFT}.
As a consequence of this ``SSFT'' in the Bjorken expansion,
we have also shown an inequality~\eqref{eq:relative-entropy-fluctuation-upper-bound}
between the initial entropy and the relative fluctuations of the final entropy.
The consequence to the experimental observables of high-energy nuclear collisions
is the inequality on the multiplicity~\eqref{eq:multiplicity-fluctuations-upper-bound}
where the left-hand side of the inequality can be directly measured in experiments,
and the right-hand side is determined solely by the initial condition models
independently from the intermediate dynamics of the system.
We also pointed out that the multiplicity fluctuations
are more significant in the smaller collision systems
as is the common nature of the thermal fluctuations.

In realistic modeling of the high-energy nuclear collisions
by the second-order causal viscous hydrodynamics,
the relaxation time is comparable to the time scale of the hydrodynamics.
In addition there would be the effects from the background fluctuations,
so we have numerically checked the breaking of the SSFT by those effects
by defining the ratio $R$~\eqref{eq:definition-of-R} from the SSFT~\eqref{eq:Bj_SSFT}.
We performed $(0+1)$-dimensional event-by-event simulations
of relativistic fluctuating hydrodynamics in the Bjorken expansion
using a stochastic Runge--Kutta method
and obtained the distribution of the final entropy production.
As a result we found that the breaking of the SSFT is more significant
for the relaxation time model that has strong temperature dependence.
To understand the result, we have analytically investigated
what effects break the SSFT with non-vanishing relaxation times
and identified three different contributions:
the finiteness of the relaxation time,
the temperature evolution and the relaxation time evolution.
The first effect vanishes in a time scale of the relaxation time,
and the second effect is relatively independent of the relaxation time model.
The third effect largely depends on the relaxation time models.
We also checked the effects of background fluctuations
by performing the event-by-event simulations
using the non-fluctuating background temperature evolution.

As a future work we are now preparing to investigate the effects
of the FDR corrections in non-static and inhomogeneous background in detail.
Also, in the present analysis, we assumed the Bjorken flow
which means that the fluctuations of the flow velocity are not considered in the analysis.
The effects of the flow fluctuations to the SSFT would be one of the future tasks.
Another interesting topic is about the definition of the entropy.
In defining the entropy production, we employed the equilibrium entropy $s$
but not the non-equilibrium entropy of the second-order hydrodynamics,
$s_{\mathrm{neq}} = s - \tau_\Pi \Pi^2/2T\zeta - \tau_\pi\pi^2/4T\eta$.
In fact the SSFT does not seem to be reproduced for the non-equilibrium entropy
in our present numerical calculations and analytical studies,
but its detailed understanding and interpretation is another future task.

%==============================================================================
\section*{Acknowledgments}
The authors thank Tomoi Koide for useful discussions.
The work of T.~H. was supported by JSPS KAKENHI Grant No.~JP17H02900.

%==============================================================================
\appendix
\section{Multiplicity fluctuations}
\label{sec:appendix-multiplicity-fluctuations}
\def\Average#1#2{\langle#2\rangle_{#1}}
\def\InitialAverage#1{\langle#1\rangle_\textrm{IS}}
\def\NoiseAverage#1{\langle#1\rangle_\xi}
\def\CooperFryeAverage#1{\langle#1\rangle_\textrm{CF}}

In this section
we derive the upper bound of the multiplicity fluctuations
in Eq.~\eqref{eq:multiplicity-fluctuations-upper-bound}.
To calculate the multiplicity fluctuations
we should distinguish three different fluctuations:
(1) initial entropy fluctuations originating from initial state fluctuations,
(2) the hydrodynamic fluctuations on which we focus in this paper
and (3) the particle number fluctuations which appear
when we switch the system description from thermodynamic fields to hadrons
using the Cooper--Frye formula~\cite{Cooper:1974mv}.
To deal with these fluctuations we define three corresponding averages:
(1) $\InitialAverage{O}$ is the average over different initial conditions,
(2) $\NoiseAverage{O}$ is the average over different noise processes
for a fixed initial entropy $\SInit$,
and (3) $\CooperFryeAverage{O}$ is the average over
particlisation by the Cooper--Frye sampling for a fixed final entropy.
The event average can be expressed as
$\EventAverage{O} = \InitialAverage{\NoiseAverage{\CooperFryeAverage{O}}}$.

With this terminology Eq.~\eqref{eq:relative-entropy-fluctuation-upper-bound}
is rewritten as
\def\STot{S_\mathrm{tot}}
\begin{align}
  \frac{\NoiseAverage{(\STot - \NoiseAverage{\STot})^2}}{\NoiseAverage{\STot}^2}
    &\le \frac{1}{2\SInit}.
\end{align}
Then we take the averages over initial conditions:
\begin{align}
  \InitialAverage{\NoiseAverage{(\STot - \NoiseAverage{\STot})^2}}
    &\le \left\langle\frac{\NoiseAverage{\STot}^2}{2\SInit}\right\rangle_\text{IS}.
  \label{eq:app-mulfluc.IS-average-upper-bound}
\end{align}
The left-hand side is decomposed into two parts as
\begin{align}
  & \InitialAverage{\NoiseAverage{(\STot - \NoiseAverage{\STot})^2}} \nonumber \\
  & \quad = \EventAverage{[(\STot - \EventAverage{\STot}) - (\NoiseAverage{\STot} - \EventAverage{\STot})]^2} \nonumber \\
  & \quad = (\Delta_\text{ev}\STot)^2 - \EventAverage{(\NoiseAverage{\STot} - \EventAverage{\STot})^2},
\end{align}
where $(\Delta_\text{ev}\STot)^2 \eqdef \EventAverage{(\STot - \EventAverage{\STot})^2}$.
We used $\EventAverage{f(\STot)} \WBR= \InitialAverage{\NoiseAverage{f(\STot)}}$
which comes from the fact that $\STot$ is independent of the particlisation,
\thatis, $\CooperFryeAverage{\STot} = \STot$.
The right-hand side is transformed as
\begin{align}
  \left\langle\frac{\NoiseAverage{\STot}^2}{2\SInit}\right\rangle_\text{IS}
  &= \left\langle\frac{\NoiseAverage{\STot}}{2}\cdot\frac{\EventAverage{\STot}}{\EventAverage{\SInit}}\right\rangle_\text{IS} \nonumber \\
  &= \frac{\EventAverage{\STot}^2}{2\EventAverage{\SInit}},
\end{align}
where we used the relation,
$\NoiseAverage{\STot}/\SInit \WBR= \EventAverage{\STot}/\EventAverage{\SInit} \WBR= \mathrm{const}$,
coming from the assumption that
both of $\NoiseAverage{\STot}$ and $\SInit$ are proportional to the transverse area $A$
of each initial condition.
Plugging them into Eq.~\eqref{eq:app-mulfluc.IS-average-upper-bound},
we obtain the upper bound of the relative fluctuations of the final entropy with initial fluctuations considered:
\begin{align}
  \frac{(\Delta_\text{ev}\STot)^2}{\EventAverage{\STot}^2}
  &\le \frac{\EventAverage{(\NoiseAverage{\STot} - \EventAverage{\STot})^2}}{\EventAverage{\STot}^2}
    + \frac1{2\EventAverage{\SInit}} \nonumber \\
  &= \frac{(\Delta_\text{ev}\SInit)^2}{\EventAverage{\SInit}^2}
    + \frac1{2\EventAverage{\SInit}}.
  \label{eq:app-mulfluc.entropy-fluctuations-upper-bound}
\end{align}
To obtain the second line
we again used the relation, $\NoiseAverage{\STot} \WBR\propto \SInit \WBR\propto A$.

Next we will relate the entropy fluctuations to the multiplicity fluctuations.
Since we assumed the Poisson distribution for the particlisation,
the mean and variance of the multiplicity for a fixed final entropy becomes
$\CooperFryeAverage{N} = \CooperFryeAverage{(N-\CooperFryeAverage{N})^2} = \alpha \STot$
with $\alpha$ being a proportionality constant.
Using this relation we obtain the multiplicity fluctuations as follows:
\begin{align}
  & \frac{\EventAverage{(N - \EventAverage{N})^2}}{\EventAverage{N}^2} \nonumber \\
  & \quad = \frac{\EventAverage{[(N - \CooperFryeAverage{N}) + (\CooperFryeAverage{N} - \EventAverage{N})]^2}}{\EventAverage{N}^2} \nonumber \\
  & \quad = \frac{\EventAverage{(N - \CooperFryeAverage{N})}}{\EventAverage{N}^2}
    + \frac{\EventAverage{(\CooperFryeAverage{N} - \EventAverage{N})^2}}{\EventAverage{N}^2} \nonumber \\
  & \quad = \frac{(\Delta_\text{ev}\STot)^2}{\EventAverage{\STot}^2}
    + \frac1{\EventAverage{N}}.
  \label{eq:app-mulfluc.multiplicity-fluctuations}
\end{align}

Combining Eqs.~\eqref{eq:app-mulfluc.entropy-fluctuations-upper-bound}~%
and~\eqref{eq:app-mulfluc.multiplicity-fluctuations},
we obtain Eq.~\eqref{eq:multiplicity-fluctuations-upper-bound}.
Finally we note that,
as the origin of the hydrodynamic fluctuations
is the microscopic degrees of freedom,
a part of the particlisation fluctuations may be already contained
in the hydrodynamic fluctuations.
Nevertheless, the inequality is still valid since
in such a case the upper bound is just overestimated.

%==============================================================================
\section{Stochastic Runge--Kutta method}
\label{sec:srk2}
In our numerical simulations we used a second-order stochastic Runge--Kutta method
for the Stratonovich stochastic differential equations.
Our equations of fluctuating hydrodynamics in one-dimensionally expanding systems
can be summarized in the following structure:
\begin{align}
  \D X_i(\tau) = f_i(\tau, \vec{X}(\tau)) + \sum_a g_{ia}(\tau,\vec{X}(\tau))\circ w_a(\tau),
\end{align}
where $\D = \d/\d\tau$ is the time derivative,
$\vec{X}(\tau) = \{X_i(\tau)\}_i$ is the set of dynamical variables,
and $\{w_a(\tau)\}_a$ are the noise terms
satisfying the normalization $\langle w_a(\tau)w_b(\tau')\rangle = \delta_{ab}\delta(\tau-\tau')$.
The coefficients, $f_i(\tau, \vec{X})$ and $g_{ia}(\tau, \vec{X})$, are
the average and fluctuating parts of the time derivatives, respectively.
In the stochastic Runge--Kutta method, which we employed,
one calculates the next-step state, $X_i^{(n+1)} = X_i(\tau_{n+1} = \tau_n + \Delta\tau)$,
from the previous-step state, $X_i^{(n)} = X_i(\tau_n)$, using the following equations:
\begin{align}
  K^{(1)}_i
    &= f_i(\tau_n, \vec{X}^{(n)})\Delta\tau \nonumber \\
    &+ \sum_a g_{ia}(\tau_n, \vec{X}^{(n)})\Delta W_a, \\
  K^{(2)}_i
    &= X^{(n)}_i + f_i(\tau_{n+1}, \vec{X}^{(n)} + \vec{K}^{(1)})\Delta\tau \nonumber \\
    &+ \sum_a g_{ia}(\tau_{n+1}, \vec{X}^{(n)} + \vec{K}^{(1)})\Delta W_a, \\
  X^{(n+1)}_i
    &= X^{(n)}_i + \frac12(K^{(1)}_{i} + K^{(2)}_{i}),
\end{align}
where $\{\Delta W_a\}_a$ are independent Gaussian random numbers of the standard deviation $\sqrt{\Delta\tau}$.

%==============================================================================
\section{Numerical integration of analytic corrections}
\label{sec:numerical-integration}
Here we describe an efficient way to evaluate the time dependence of
the analytic corrections $\gamma$, $\delta_t$ and $\delta_r$.
First we transform the expressions as follows:
\begin{align}
  \gamma(\tau) &= -\frac{\cellVolume}{2(\tau-\tau\init)}
      \frac{\tauRz(\tau)}{T_0(\tau)} I(\tau), \\
  \delta_a(\tau) &=
      \frac{\cellVolume}{2(\tau-\tau\init)} \Delta_a(\tau), \\
  \Delta_a(\tau) &=
      \int_{\tau\init}^\tau \frac{\d\tau_1}{T_0(\tau_1)} I_a(\tau_1)
      +\int_{\tau\init}^\tau \d\tau_3 D_a(\tau_3) I(\tau_3), \\
  I(\tau) &= \int_{\tau\init}^\tau \d\tau_1
      G_0(\tau, \tau_1) \frac{\tauRz(\tau_1)}{T_0(\tau_1)}
      \Gamma(\tau_1), \\
  I_a(\tau) &= \int_{\tau\init}^{\tau} \d\tau_3
      G_0(\tau, \tau_3)\tauRz(\tau_3) D_a(\tau_3) \Gamma(\tau_3), \\
  \Gamma(\tau) &= \int_{\tau\init}^\tau \d\tau_1
      G_{1/2}(\tau, \tau_1)\frac{T_0(\tau_1)}{\tauRz(\tau_1)}\langle F(\tau_1)\rangle,
\end{align}
where the subscript $a$ is either $t$ or $r$,
$D_t(\tau)\WBR = \WBR\tauRz(\tau)\WBR\D\WBR(1/T_0(\tau))$ and $D_r(\tau) = (1/T_0(\tau))\D\tauRz(\tau)$.
The function $G_{1/2}(\tau, \tau_1)$ is
the Green function of the ``half relaxation time'' defined as
\begin{align}
  G_{1/2}(\tau, \tau_1)
    &= 2G_0(\tau,\tau_1)^2 \tauRz(\tau_1) \nonumber \\
    &= \exp\Bigl(-\int_{\tau_1}^{\tau} \frac{\d\tau'}{\tauRz(\tau')/2}\Bigr) \frac1{\tauRz(\tau_1)/2}.
\end{align}

Then we find the following differential equations
for the above integrations:
\begin{align}
  \left[1+\frac{\tauRz(\tau)}2 \D\right] \Gamma(\tau) &= \frac{T_0(\tau)}{\tauRz(\tau)}\langle F(\tau)\rangle, \\
  [1 + \tauRz(\tau)\D] I(\tau) &= \frac{\tauRz(\tau)}{T_0(\tau)} \Gamma(\tau), \\
  [1 + \tauRz(\tau)\D] I_a(\tau) &= \tauRz(\tau) D_a(\tau) \Gamma(\tau), \\
  \D\Delta_a &= \frac1{T_0(\tau)} I_a(\tau) + D_a(\tau) I(\tau).
\end{align}
Starting from the initial conditions
$\Gamma(\tau\init) = I(\tau\init) = I_a(\tau\init) = \Delta_a(\tau\init) = 0$,
one can solve these equations to obtain the time dependence of corrections
using such as the Runge--Kutta methods.

%==============================================================================

%\bibliographystyle{utphys}
%\bibliographystyle{apsrev4-1}
\bibliographystyle{elsarticle-num}
\bibliography{fluctuation}

\end{document}